\documentclass{article}
\usepackage{gsirep}
\usepackage{epsfig}
\title{Universal Pion Freeze-out Phase-Space Density}
\author{D. Ferenc, B. Tom\'a\v sik, U. Heinz, 
Inst. f. Theor. Physik, Universit\"at Regensburg
         }
\date{}
\begin{document}
\maketitle

\noindent
G. Bertsch has indicated~\cite{Bertsch} a possibility to 
measure the pion freeze-out phase-space density 
and thereby test the local thermal equilibrium in a
pion source. In case of thermal equilibrium at temperature
$T$,
identical pions of energy $E$ would follow
the Bose-Einstein distribution 

\begin{equation}
 \label{16}
f = \frac {1}{e^\frac{E}{T}-1}.
\end{equation}

\noindent
An average of this function over different
phase-space regions is the quantity to be measured.
%
%
\noindent
When the
$p_{T}$-spectrum is parameterized by an exponential
with $T_{\rm eff}(y)$ being the inverse slope parameter,
averaging over the spatial 
coordinates yields

\begin{equation}
\label{11}
 \langle f \rangle (p_{T},y) = \frac
{ { \frac{\sqrt{\pi}}{2}\,
 \frac{\sqrt{\lambda_{\rm dir}(p_{T},y)}}{E_p \, T^2_{\rm eff}(y)}\,
 \exp  \left (-\frac{p_{T}}{T_{\rm eff}(y)} \right )\, \frac{dn^-}{dy}(y) } }
{{R_s(p_{T},y) \sqrt{R^2_o(p_{T},y) R^2_l(p_{T},y) - R^4_{ol}(p_{T},y
)}}
 \,}.
 \end{equation}


\noindent
This equation comprises information from essentially two different
classes of experimental results:
the single particle momentum spectra ($dn^-/dy, T_{eff}$), and
the two pion Bose-Einstein correlations 
($R_s, R_o, R_l, R_{ol}, \lambda_{dir}$).
We have calculated $\langle f \rangle (p_{T},y)$ for the
S-S, S-Cu, S-Ag, S-Au, S-Pb and Pb-Pb data from
the experiments 
NA35~\cite{NA35}, NA49~\cite{NA49}, NA44~\cite{NA44}, 
and for the $\pi$-p data from the NA22
experiment~\cite{NA22} at CERN-SPS. 
\begin{figure}[htb]
\epsfig{file=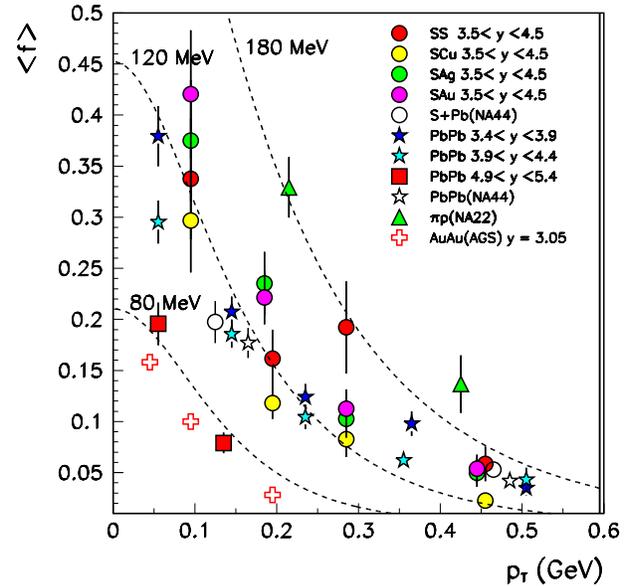,width=9.0cm}
\caption{
\noindent
Phase-space density as a function of $p_T$ for different data sets.
Heavy-ion data from SPS are indistinguishably similar, although they
span over an order of magnitude range in multiplicity.
A Bose-Einstein function (Eq.~(\ref{16})) is superimposed with the
three choices of the local freeze-out 
temperature: 80 MeV, 120 MeV and 180 MeV. 
}
\label{FPP2}
\end{figure}
\noindent
From the results for $\langle f \rangle$ as a function
of $p_T$, presented in Fig.~\ref{FPP2}, 
one may conclude:{\vskip 0.15 cm}

\noindent
{\bf 1. universal phase-space density}\\
All the nuclear collision data 
from the SPS in Fig.~\ref{FPP2} are indistinguishably similar,
in spite of a factor of $\sim$ 10 difference 
in multiplicity density.{\vskip 0.15 cm}

\noindent
{\bf 2. agreement with Bose-Einstein distribution}\\
Using simultaneously pion spectra and pion correlations
NA49 has disentangled thermal
motion from collective expansion~\cite{NA49}.
Taking the NA49 result for the local freeze-out temperature
$T$ = 120 MeV as 
the {\bf only} parameter in Eq.~(\ref{16})
one indeed finds good agreement with the data.
This is consistent with the thermal nature of
the pion source, and inconsistent with
the presence of a hypothetic pion condensate.{\vskip 0.15 cm}

\noindent
{\bf 3. radial flow}\\
Looking in more detail, one finds that
the data indicate a somewhat slower decrease with increasing $p_T$
than the Bose-Einstein curve.
This is most likely due to radial collective expansion which
adds extra transverse momentum to particles, i.e. the 
local $\langle f \rangle$ values appear in the measurement at a $p_T$
that is higher than the local $\langle p_{T}\rangle$ in the source 
reference frame. A detailed study is under way.{\vskip 0.15 cm}

\noindent
{\bf 4. rapidity dependence}\\
A certain  departure 
from the universal scaling is seen
for the data at rapidities close to the 
projectile rapidity, both 
at AGS and SPS; moreover, the two results are consistent.{\vskip 0.15 cm}

\noindent
{\bf 5. high temperature decoupling in $\pi-p$ collisions}\\
In contrast to freeze-out in nuclear collisions
which takes place in two steps (chemical at $T \simeq$ 170-180 MeV 
and thermal at $T \simeq$ 
120 MeV),
pion production in $\pi$-p collisions~\cite{NA22} is essentially
immediate, without the second evolution stage, and therefore freeze-out
temperatures of around 180 MeV should be 
expected. The data~\cite{NA22} are indeed consistent with this expectation,
as seen in Fig.~\ref{FPP2}.

\end{document}